# A Prospective Randomized Clinical Trial for Measuring Radiology Study Reporting Time on Artificial Intelligence-Based Detection of Intracranial Hemorrhage in Emergent Care Head CT

Axel Wismüller[1,2,3,4] and Larry Stockmaster[1]

[1]Department of Imaging Sciences, University of Rochester Medical Center, Rochester, NY, USA
[2]Department of Biomedical Engineering, University of Rochester, NY, USA
[3]Department of Electrical Engineering, University of Rochester, NY, USA
[4]Faculty of Medicine and Institute of Clinical Radiology, Ludwig Maximilian University, Munich, Germany

## ABSTRACT

The quantitative evaluation of Artificial Intelligence (AI) systems in a clinical context is a challenging endeavor, where the development and implementation of meaningful performance metrics is still in its infancy. Here, we propose a scientific concept, Artificial Intelligence Prospective Randomized Observer Blinding Evaluation (AI-PROBE) for quantitative clinical performance evaluation of radiology AI systems within prospective randomized clinical trials. Our evaluation workflow encompasses a study design and a corresponding radiology Information Technology (IT) infrastructure that randomly blinds radiologists with regards to the presence of positive reads as provided by AI-based image analysis systems. To demonstrate the applicability of our AI-evaluation framework, we present a first prospective randomized clinical trial on investigating the effect of automatic identification of Intra-Cranial Hemorrhage (ICH) in emergent care head CT scans on radiology study Turn-Around Time (TAT) in a clinical environment. Here, we acquired 620 consecutive non-contrast head CT scans from CT scanners used for inpatient and emergency room patients at a large academic hospital over a time period of 14 consecutive days. Immediately following image acquisition, scans were automatically analyzed for the presence of ICH using commercially available software (Aidoc, Tel Aviv, Israel). Cases identified as positive for ICH by AI (ICH-AI+) were automatically flagged in the radiologists' reading worklists, where flagging was randomly switched off with a probability of 50%. Study TAT was measured automatically as the time difference between study completion and first clinically communicated study reporting, with time stamps for these events automatically retrieved from various radiology IT systems. TATs for flagged cases (73 ± 143 min) were significantly lower than TATs for non-flagged (132 ± 193 min) cases ($p<0.05$, one-sided $t$-test), where 105 of the 122 ICH-AI+ cases were true positive reads. Total sensitivity, specificity, and accuracy over all analyzed cases were 95.0%, 96.7%, and 96.4%, respectively. We conclude that automatic identification of ICH reduces study TAT for ICH in emergent care head CT settings, which carries the potential for improving clinical management of ICH by accelerating clinically indicated therapeutic interventions. In a broader context, our results suggest that our AI-PROBE framework can contribute to a systematic quantitative evaluation of AI systems in a clinical workflow environment with regards to clinically meaningful performance measures, such as TAT or diagnostic accuracy metrics.

## 1. INTRODUCTION

A key challenge for the adoption of Artificial Intelligence (AI) systems in a clinical context is the generic lack of standards for objectively evaluating such systems with regards to clinically meaningful performance metrics. Solely relying on classical performance metrics from machine learning or signal processing, such as Receiver Operating Characteristic (ROC) analysis of diagnostic accuracy or computation times, may be helpful for initially evaluating and comparing AI systems on benchmark data in isolation from their actual clinical application context, but they fall short at evaluating the clinical usefulness of AI systems in real-world settings.

A fully automated diagnosis by AI systems without human intervention would potentially require a 'strong' artificial intelligence to be socially acceptable. As artificial intelligence, despite its 70-year history, is still in its infancy, the deployment and operation of AI systems in medicine will continue to require human control and supervision in the foreseeable future. Hence, an objective evaluation of contemporary AI systems can only be based on the interplay of such systems with human experts using them in a real-world context, such as radiologists in a clinical reading workflow environment. Here, it is important to understand that not only the performance of the 'inference machine' in AI systems, such as their classification or regression quality, determines their real-world usefulness, but that other factors, such as user interface, ease of access, user motivation, etc., may play a critical role as well.

Although most of the academic literature on AI in medical imaging focuses on comparing the performance of different AI systems on given data, such as in various 'competitions' on publicly available benchmark data, see e.g. [1], [2], [3], it is often more important to investigate the effects of deployment of a *single* AI system based on quantitative measurable clinical parameters. The reason for this is that AI systems, such as other capital investments in the healthcare enterprise, are usually not introduced in a clinical workflow environment for the sake of comparing them to each other. Rather a decision is made for deploying a single system in a given environment, based on administrative, financial or business development considerations. In radiology, for example, the challenge of evaluating the clinical usefulness of an AI system would be to compare the radiologists' performance with and without using the system, where clinically meaningful performance measures are determined, such as diagnostic accuracy or study turnaround time, resulting from human-machine inter-operability.

To address this challenge, we propose a scientific framework, Artificial Intelligence Prospective Randomized Observer Blinding Evaluation (AI-PROBE) for quantitative clinical performance evaluation of radiology AI systems within prospective randomized clinical trials. Our evaluation workflow encompasses a study design and a corresponding radiology Information Technology (IT) infrastructure that randomly blinds radiologists with regards to the presence of positive reads as provided by AI-based image analysis systems.

In the following, we will explain the basic idea of AI-PROBE, describe its clinical application within a prospective randomized clinical trial for AI-based detection of ICH in emergent care head CT scans, and will report study results on radiology turnaround times for this clinical application. This work is embedded in our group's endeavor to expedite artificial intelligence in biomedical imaging by means of advanced pattern recognition and machine learning methods for computational radiology and radiomics, e.g. [4–54].

## 2. SCIENTIFIC CONCEPT

### 2.1 Shortcomings of Retrospective AI evaluation

The potential of retrospective performance analysis of radiology AI systems is limited and subject to manifold systematic errors. A typical scenario for retrospective analysis would be to compare certain clinical outcome measures before and after deployment of an AI system into a radiology workflow. Such systematic errors include temporal changes before and after AI deployment, such as changing radiology staffing, work organization patterns, or temporal shifts of case distributions, to name a few. In order to avoid such systematic errors, it follows that retrospective AI evaluation approaches, although widely used, should be replaced by prospective evaluation methods. However, introducing prospective radiology AI evaluation is challenging, because (i) there is limited experience and literature on prospective approaches to radiology AI system evaluation, (ii) such approaches imply significant both organizational and technical efforts.

### 2.2 Artificial Intelligence Prospective Randomized Observer Blinding Evaluation (AI-PROBE)

To address this challenge, we propose a scientific concept, Artificial Intelligence Prospective Randomized Observer Blinding Evaluation (AI-PROBE), for quantitative clinical performance evaluation of radiology AI systems within prospective randomized clinical trials. Our evaluation workflow encompasses a study design that randomly blinds radiologists with regards to the presence of reading results provided by AI-based image analysis systems. There are various options within this concept, which can be chosen according to the clinical needs of a specific evaluation situation.

(i) The first option, AI-PROBE-1, is specifically suited for radiology worklist prioritization, where radiologists may only be interested to be informed about positive reads, for example, flagging of cases, in which an AI system has identified an emergent finding, such as an intracranial hemorrhage on a head CT scan. In this scenario, having no flag associated with a radiology study does not provide any information to the observer, as to whether she was blinded about the AI reading results based on the AI-PROBE clinical trial design, whether the study was analyzed at all, or whether it was analyzed, but has been rated by the AI system as a negative read for the examined condition.

**Figure 1:** Box and whiskers plots for study turnaround times for flagged and non-flagged CT cases identified as positive for intracranial hemorrhage by a commercial image analysis system. Note that non-flagged cases appear to have a higher turnaround time (one-sided *t*-test) with means of 73 min for flagged cases, and 132 min for non-flagged cases. Horizontal lines represent the $25^{th}$, $50^{th}$, and $75^{th}$ percentiles, respectively. For details, see text.

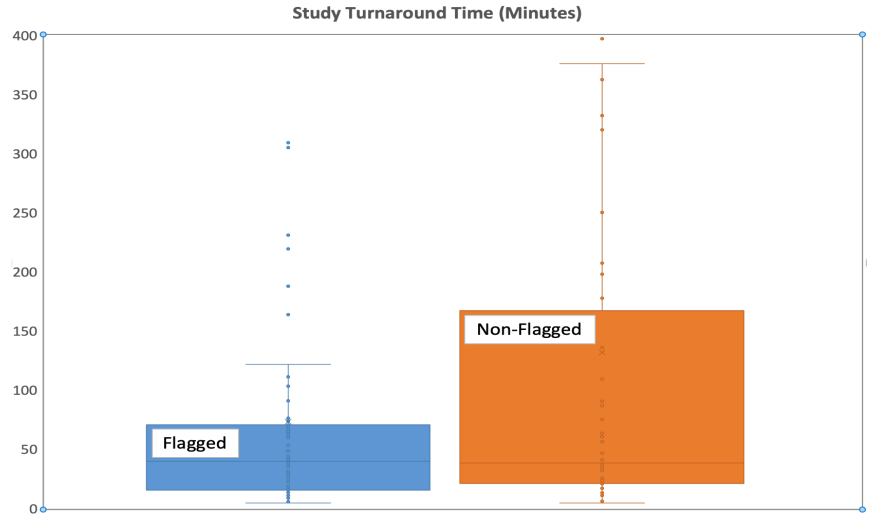

(ii) The second option, AI-PROBE-2, randomly blinds the observer about the results of AI system reads, but flags all cases that have been analyzed by the AI system. In this scenario, different types of flags need to be utilized. In the simplest case, the user would have to be informed, whether a case was read as positive or negative by the AI system for an examined condition, such as for the presence or absence of an intracranial hemorrhage. Having no flag associated with a given radiology study informs the user that there are no AI reading results available for this case for any reason, be it that the user was blinded for the given case according to the AI-PROBE clinical trial design, or that there are no AI reading results available for any technical reasons. Such technical reasons can be manifold, including (i) that a given scanner or imaging archive was not connected to the AI system, (ii) the case was not sent for analysis, (iii) the case was sent, but not analyzed, (iv) the case was analyzed, but no results are available, such as induced by network communication or inference machine processing delays. Note that this list of reasons is not exhaustive, and that an arbitrary number of additional flags may be introduced to inform the user about such reasons.

A free parameter that needs to be considered is the probability $P$ for selecting cases, for which radiologists will be blinded with regards to AI reading results. For simplicity, in the trial on intracranial hemorrhage reported below, we chose an AI-PROBE-1 study design with $P=50\%$.

We emphasize that multiple organizational and technical challenges have to be met simultaneously for implementing an AI-PROBE study design. These challenges include, but are not limited to, (i) acquiring Institutional Review Board approval for radiologists' participation in such prospective trials, (ii) aligning the interests of diverse stakeholders, including radiologists, hospital administrators, and AI solution vendors, and (iii) significant interface programming efforts for automatically retrieving heterogeneous information from multiple radiology and hospital IT systems, such as various time stamps for events related to technical processing or human intervention.

## 3. CLINICAL APPLICATION: INTRACRANIAL HEMORRHAGE DETECTION TRIAL

**Clinical Relevance:** To demonstrate the applicability of our AI-evaluation framework, we present a first prospective randomized trial on investigating the effect of automatic identification of Intra-Cranial Hemorrhage (ICH) in emergent care head CT scans on radiology study Turn-Around Time (TAT) in a clinical environment. Timely detection of ICH on medical imaging studies is critical, because delayed therapeutic interventions in emergency settings may be detrimental for patient outcome. For example, in hemorrhagic stroke, the American Heart Association (AHA)/American Stroke Association (ASA) 2018 guidelines for early management emphasize the time-dependent benefit of tissue plasminogen activator (tPA) therapy [55]. Patients with ischemic stroke must be identified as soon as possible so that tPA can be administered within 3-4.5 hours of symptom onset [55]. A non-contrast head CT scan, which must be negative for hemorrhage, is a required diagnostic test before tPA can be administered in a timely manner [56,57,58]. The clinical outcomes of patients receiving tPA within a designated time limit are closely linked to time of treatment [58,59]. Besides hemorrhagic stroke, there are numerous other clinical conditions, in which ICH will lead to rapid increase of intracranial pressure, where early detection and treatment of ICH can significantly reduce patient morbidity and mortality.

**Data and Methods:** A total of 620 consecutive non-contrast head CT scans from two CT scanners used for inpatient and emergency room patients at a large academic hospital (Strong Memorial Hospital, University of Rochester Medical Center) were prospectively acquired over a time period of 14 consecutive days. Immediately following image acquisition, scans were automatically analyzed for the presence of intracranial hemorrhage (ICH) using commercially available software (Aidoc, Tel Aviv, Israel). Cases identified as positive for ICH by AI (ICH-AI+) were automatically flagged in the radiologists' reading worklists, where flagging was randomly switched off with a probability $P$=50%. Study turnaround time (TAT) was measured automatically as the time difference between study completion time (=study accessible to radiologists for reporting) to study reporting time (=first report visible to clinicians, regardless whether preliminary or final). Time stamps for calculating TAT were automatically retrieved from various radiology IT systems. The hypothesis that TATs for non-flagged ICH-AI+ cases were higher than for flagged ICH-AI+ cases was tested using a one-sided heteroscedastic $t$-test. Total sensitivity, specificity, and accuracy for ICH detection were calculated for all analyzed 620 cases, where final radiology reports served as ground truth.

## 4. RESULTS

A total of 122 ICH-AI+ cases were found among the total number of analyzed 620 non-contrast head CT cases, of which 66 cases were flagged. Study TAT was measured automatically as the time difference between study completion time, i.e., when the study became accessible to radiologists for reporting, and the study reporting time, i.e., when the first report was completed that was visible to clinicians, regardless whether this report legally qualified as a preliminary or as a final report. Figure 1 summarizes our results on the distributions of radiology study turnaround times (TAT), comparing flagged and non-flagged ICH-AI+ cases. It can be seen that the TAT distributions for flagged and non-flagged cases are different, with higher TATs observed for non-flagged ICH-AI+ cases. This is confirmed by statistical analysis: TATs for flagged cases (73 ± 143 min) were significantly lower than TATs for non-flagged (132 ± 193 min) cases ($p$<0.05, one-sided $t$-test). A total of 105 of the 122 ICH-AI+ cases were true positive reads, as verified by human review of final radiology reports. Total sensitivity, specificity, and accuracy over all analyzed 620 cases were 95.0%, 96.7%, and 96.4%, respectively.

## 5. NEW AND BREAKTHROUGH WORK

We introduce a scientific framework, Artificial Intelligence Prospective Randomized Observer Blinding Evaluation (AI-PROBE) for quantitative clinical performance evaluation of radiology AI systems within prospective randomized clinical trials. Our evaluation workflow encompasses a study design and a corresponding radiology information technology infrastructure that randomly blinds radiologists with regards to the presence of positive reads as provided by AI-based image analysis systems. To demonstrate the applicability of our AI-evaluation framework, we present a first prospective randomized clinical trial on investigating the effect of automatic identification of Intra-Cranial Hemorrhage (ICH) in 620 emergent care head CT scans on radiology study Turn-Around Time (TAT) in a clinical environment. Our results suggest that notifying radiologists on automatically detected ICH reduces TAT for reporting ICH to clinicians in emergency setting head CT scans, as shown by our prospective, randomized clinical study. Such reduced TAT may expedite clinically indicated therapeutic interventions.

## 6. CONCLUSION

We conclude from our prospective randomized clinical trial that automatic identification of Intra-Cranial Hemorrhage (ICH) reduces radiology study turnaround times for ICH in emergent care head CT settings, which carries the potential for improving clinical management of ICH by accelerating clinically indicated therapeutic interventions. In a broader scientific context, our results suggest that the proposed AI-PROBE framework can contribute to a systematic quantitative evaluation of AI systems in a real-world workflow environment with regards to clinically meaningful performance metrics, such as radiology study turnaround times or diagnostic accuracy measures.

## ACKNOWLEDGEMENTS

This research was funded by a Ernest J. Del Monte Institute for Neuroscience Award from the Harry T. Mangurian Jr. Foundation. This work was conducted as a Practice Quality Improvement (PQI) project related to American Board of Radiology (ABR) Maintenance of Certificate (MOC) for Prof. Dr. Axel Wismüller. This work is not being and has not been submitted for publication or presentation elsewhere. The authors are grateful to Drs. E. Weinberg and K. Chugthai, University of Rochester Medical Center, and Shuli Edwards, Aidoc Inc., for their support.

# REFERENCES

[1] Luciano M. Prevedello, Safwan S. Halabi, George Shih, Carol C. Wu, Marc D. Kohli, Falgun H. Chokshi, Bradley J. Erickson, Jayashree Kalpathy-Cramer, Katherine P. Andriole, and Adam E. Flanders: Challenges Related to Artificial Intelligence Research in Medical Imaging and the Importance of Image Analysis Competitions. Radiology: Artificial Intelligence 2019 1:1

[2] Justin Ker, Lipo Wang, Jai Rao, and Tchoyoson Lim. Deep Learning Applications in Medical Image Analysis. IEEE Access 6:9375-9389 (2018)

[3] https://grand-challenge.org/challenges/

[4] Nattkemper, T. W. and Wismüller, A., "Tumor feature visualization with unsupervised learning," Medical Image Analysis 9(4), 344–351 (2005).

[5] Bunte, K., Hammer, B., Wismüller, A., and Biehl, M., "Adaptive local dissimilarity measures for discriminative dimension reduction of labeled data," Neurocomputing 73(7-9), 1074–1092 (2010).

[6] Wismüller, A., Vietze, F., and Dersch, D. R., "Segmentation with neural networks," in [Handbook of Medical Imaging], 107–126, Academic Press, Inc. (2000).

[7] Leinsinger, G., Schlossbauer, T., Scherr, M., Lange, O., Reiser, M., and Wismüller, A., "Cluster analysis of signal intensity time course in dynamic breast MRI: does unsupervised vector quantization help to evaluate small mammographic lesions?," European Radiology 16(5), 1138–1146 (2006).

[8] Wismüller, A., Vietze, F., Behrends, J., Meyer-Bäse, A., Reiser, M., and Ritter, H., "Fully automated biomedical image segmentation by self-organized model adaptation," Neural Networks 17(8-9), 1327–1344 (2004).

[9] Hoole, P., Wismüller, A., Leinsinger, G., Kroos, C., Geumann, A., and Inoue, M., "Analysis of tongue configuration in multi-speaker, multi-volume MRI data,"

[10] Wismüller, A., "Exploratory Morphogenesis (XOM): a novel computational framework for self-organization," Ph. D. thesis, Technical University of Munich, Department of Electrical and Computer Engineering (2006).

[11] Wismüller, A., Dersch, D. R., Lipinski, B., Hahn, K., and Auer, D., "A neural network approach to functional MRI pattern analysis—clustering of time-series by hierarchical vector quantization," in [International Conference on Artificial Neural Networks], 857–862, Springer, London (1998).

[12] Wismüller, A., Vietze, F., Dersch, D. R., Behrends, J., Hahn, K., and Ritter, H., "The deformable feature map – a novel neurocomputing algorithm for adaptive plasticity in pattern analysis," Neurocomputing 48(1-4), 107–139 (2002).

[13] Behrends, J., Hoole, P., Leinsinger, G. L., Tillmann, H. G., Hahn, K., Reiser, M., and Wismüller, A., "A segmentation and analysis method for MRI data of the human vocal tract," in [Bildverarbeitung für die Medizin 2003], 186–190, Springer, Berlin, Heidelberg (2003).

[14] Wismüller, A., "Neural network computation in biomedical research: chances for conceptual cross-fertilization," Theory in Biosciences (1997).

[15] Bunte, K., Hammer, B., Villmann, T., Biehl, M., and Wismüller, A., "Exploratory observation machine (XOM) with Kullback-Leibler divergence for dimensionality reduction and visualization.," in [ESANN], 10, 87–92 (2010).

[16] Wismüller, A., Vietze, F., Dersch, D. R., Hahn, K., and Ritter, H., "The deformable feature map—adaptive plasticity for function approximation," in [International Conference on Artificial Neural Networks], 123–128, Springer, London (1998).

[17] Wismüller, A., "The exploration machine–a novel method for data visualization," in [International Workshop on Self-Organizing Maps], 344–352, Springer, Berlin, Heidelberg (2009).

[18] Wismüller, A., "Method, data processing device and computer program product for processing data," (July 28 2009). US Patent 7,567,889.

[19] Meyer-Bäse, A., Jancke, K., Wismüller, A., Foo, S., and Martinetz, T., "Medical image compression using topology preserving neural networks," Engineering Applications of Artificial Intelligence 18(4), 383–392 (2005).

[20] Huber, M. B., Nagarajan, M., Leinsinger, G., Ray, L., and Wismüller, A., "Classification of interstitial lung disease patterns with topological texture features," in [Medical Imaging 2010: Computer-Aided Diagnosis], 7624, 762410, International Society for Optics and Photonics (2010).

[21] Wismüller, A., "The exploration machine: a novel method for analyzing high-dimensional data in computer-aided diagnosis," in [Medical Imaging 2009: Computer-Aided Diagnosis], 7260, 72600G, International Society for Optics and Photonics (2009).

[22] Bunte, K., Hammer, B., Villmann, T., Biehl, M., and Wismüller, A., "Neighbor embedding XOM for dimension reduction and visualization," Neurocomputing 74(9), 1340–1350 (2011).

[23] Meyer-Bäse, A., Lange, O., Wismüller, A., and Ritter, H., "Model-free functional MRI analysis using topographic independent component analysis," International Journal of Neural Systems 14(04), 217–228 (2004).

[24] Wismüller, A., "A computational framework for nonlinear dimensionality reduction and clustering," in [International Workshop on Self-Organizing Maps], 334–343, Springer, Berlin, Heidelberg (2009).

[25] Meyer-Bäse, A., Auer, D., and Wismüller, A., "Topographic independent component analysis for fMRI signal detection," in [Proceedings of the International Joint Conference on Neural Networks, 2003.], 1, 601–605, IEEE (2003).

[26] Meyer-Bäse, A., Schlossbauer, T., Lange, O., and Wismüller, A., "Small lesions evaluation based on unsupervised cluster analysis of signal-intensity time courses in dynamic breast MRI," Journal of Biomedical Imaging 2009, 31 (2009).

[27] Wismüller, A., Lange, O., Auer, D., and Leinsinger, G., "Model-free functional MRI analysis for detecting low-frequency

functional connectivity in the human brain," in [Medical Imaging 2010: Computer-Aided Diagnosis], 7624, 76241M, International Society for Optics and Photonics (2010).

[28] Huber, M. B., Nagarajan, M. B., Leinsinger, G., Eibel, R., Ray, L. A., and Wismüller, A., "Performance of topological texture features to classify fibrotic interstitial lung disease patterns," Medical Physics 38(4), 2035–2044 (2011).

[29] Wismüller, A., Verleysen, M., Aupetit, M., and Lee, J. A., "Recent advances in nonlinear dimensionality reduction, manifold and topological learning.," in [ESANN], (2010).

[30] Wismüller, A., Behrends, J., Hoole, P., Leinsinger, G. L., Reiser, M. F., and Westesson, P.-L., "Human vocal tract analysis by in vivo 3D MRI during phonation: a complete system for imaging, quantitative modeling, and speech synthesis," in [International Conference on Medical Image Computing and Computer-Assisted Intervention], 306–312, Springer, Berlin, Heidelberg (2008).

[31] Huber, M. B., Bunte, K., Nagarajan, M. B., Biehl, M., Ray, L. A., and Wismüller, A., "Texture feature ranking with relevance learning to classify interstitial lung disease patterns," Artificial Intelligence in Medicine 56(2), 91–97 (2012).

[32] Wismüller, A., Meyer-Bäse, A., Lange, O., Reiser, M., and Leinsinger, G., "Cluster analysis of dynamic cerebral contrast-enhanced perfusion MRI time-series," Medical Imaging, IEEE Transactions on 25(1), 62–73 (2006).

[33] Twellmann, T., Saalbach, A., Müller, C., Nattkemper, T. W., and Wismüller, A., "Detection of suspicious lesions in dynamic contrast enhanced MRI data," in [The 26th Annual International Conference of the IEEE Engineering in Medicine and Biology Society], 1, 454–457, IEEE (2004).

[34] Schlossbauer, T., Leinsinger, G., Wismüller, A., Lange, O., Scherr, M., Meyer-Bäse, A., and Reiser, M., "Classification of small contrast enhancing breast lesions in dynamic magnetic resonance imaging using a combination of morphological criteria and dynamic analysis based on unsupervised vector-quantization," Investigative Radiology 43(1), 56 (2008).

[35] Otto, T. D., Meyer-Bäse, A., Hurdal, M., Sumners, D., Auer, D., and Wismüller, A., "Model-free functional MRI analysis using cluster-based methods," in [Intelligent Computing: Theory and Applications], 5103, 17–24, International Society for Optics and Photonics (2003).

[36] Varini, C., Nattkemper, T. W., Degenhard, A., and Wismüller, A., "Breast MRI data analysis by LLE," in [2004 IEEE International Joint Conference on Neural Networks (IEEE Cat. No. 04CH37541)], 3, 2449–2454, IEEE (2004).

[37] Huber, M. B., Lancianese, S. L., Nagarajan, M. B., Ikpot, I. Z., Lerner, A. L., and Wismüller, A., "Prediction of biomechanical properties of trabecular bone in MR images with geometric features and support vector regression," IEEE Transactions on Biomedical Engineering 58(6), 1820–1826 (2011).

[38] Meyer-Bäse, A., Pilyugin, S. S., and Wismüller, A., "Stability analysis of a self-organizing neural network with feed-forward and feedback dynamics," in [2004 IEEE International Joint Conference on Neural Networks (IEEE Cat.No. 04CH37541)], 2, 1505–1509, IEEE (2004).

[39] Meyer-Bäse, A., Lange, O., Schlossbauer, T., and Wismüller, A., "Computer-aided diagnosis and visualization based on clustering and independent component analysis for breast MRI," in [2008 15th IEEE International Conference on Image Processing], 3000–3003, IEEE (2008).

[40] Wismüller, A., Meyer-Bäse, A., Lange, O., Schlossbauer, T., Kallergi, M., Reiser, M., and Leinsinger, G., "Segmentation and classification of dynamic breast magnetic resonance image data," Journal of Electronic Imaging 15(1), 013020 (2006).

[41] Nagarajan, M. B., Huber, M. B., Schlossbauer, T., Leinsinger, G., Krol, A., and Wismüller, A., "Classification of small lesions in dynamic breast MRI: eliminating the need for precise lesion segmentation through spatio-temporal analysis of contrast enhancement," Machine Vision and Applications 24(7), 1371–1381 (2013).

[42] Nagarajan, M. B., Huber, M. B., Schlossbauer, T., Leinsinger, G., Krol, A., and Wismüller, A., "Classification of small lesions in breast MRI: evaluating the role of dynamically extracted texture features through feature selection," Journal of Medical and Biological Engineering 33(1) (2013).

[43] Bhole, C., Pal, C., Rim, D., and Wismüller, A., "3D segmentation of abdominal CT imagery with graphical models, conditional random fields and learning," Machine Vision and Applications 25(2), 301–325 (2014).

[44] Nagarajan, M. B., Coan, P., Huber, M. B., Diemoz, P. C., Glaser, C., and Wismüller, A., "Computer-aided diagnosis in phase contrast imaging x-ray computed tomography for quantitative characterization of ex vivo human patellar cartilage," IEEE Transactions on Biomedical Engineering 60(10), 2896–2903 (2013).

[45] Wismüller, A., Meyer-Bäse, A., Lange, O., Auer, D., Reiser, M. F., and Sumners, D., "Model-free functional MRI analysis based on unsupervised clustering," Journal of Biomedical Informatics 37(1), 10–18 (2004).

[46] Meyer-Bäse, A., Wismüller, A., Lange, O., and Leinsinger, G., "Computer-aided diagnosis in breast MRI based on unsupervised clustering techniques," in [Intelligent Computing: Theory and Applications II], 5421, 29–37, International Society for Optics and Photonics (2004).

[47] Nagarajan, M. B., Coan, P., Huber, M. B., Diemoz, P. C., Glaser, C., and Wismüller, A., "Computer-aided diagnosis for phase-contrast x-ray computed tomography: quantitative characterization of human patellar cartilage with high-dimensional geometric features," Journal of Digital Imaging 27(1), 98–107 (2014).

[48] Nagarajan, M. B., Huber, M. B., Schlossbauer, T., Leinsinger, G., Krol, A., and Wismüller, A., "Classification of small lesions on dynamic breast MRI: Integrating dimension reduction and out-of-sample extension into CADx methodology," Artificial Intelligence in Medicine 60(1), 65–77 (2014).

[49] Wismüller, A., Wang, X., DSouza, A. M., and Nagarajan, M. B., "A framework for exploring non-linear functional connectivity and causality in the human brain: mutual connectivity analysis (MCA) of resting-state functional MRI with convergent cross-mapping and non-metric clustering," arXiv preprint arXiv:1407.3809 (2014).

[50] Schmidt, C., Pester, B., Schmid-Hertel, N., Witte, H., Wismüller, A., and Leistritz, L., "A multivariate Granger causality concept towards full brain functional connectivity," PloS one 11(4), e0153105 (2016).

[51] DSouza, A. M., Abidin, A. Z., Leistritz, L., and Wismüller, A., "Exploring connectivity with large-scale Granger causality on resting-state functional MRI," Journal of Neuroscience Methods 287, 68–79 (2017).

[52] Chen, L., Wu, Y., DSouza, A. M., Abidin, A. Z., Wismüller, A., and Xu, C., "MRI tumor segmentation with densely connected 3D CNN," in [Medical Imaging 2018: Image Processing], 10574, 105741F, International Society for Optics and Photonics (2018).

[53] Abidin, A. Z., DSouza, A. M., Nagarajan, M. B., Wang, L., Qiu, X., Schifitto, G., and Wismüller, A., "Alteration of brain network topology in HIV-associated neurocognitive disorder: A novel functional connectivity perspective," NeuroImage: Clinical 17, 768–777 (2018).

[54] Abidin, A. Z., Deng, B., DSouza, A. M., Nagarajan, M. B., Coan, P., and Wismüller, A., "Deep transfer learning for characterizing chondrocyte patterns in phase contrast x-ray computed tomography images of the human patellar cartilage," Computers in Biology and Medicine 95, 24–33 (2018).

[55] Powers WJ, Rabinstein AA, Ackerson T, et al. 2018 Guidelines for the Early Management of Patients With Acute Ischemic Stroke: A Guideline for Healthcare Professionals From the American Heart Association/American Stroke Association. Stroke 2018;49:e46 – e110.

[56] Lees KR, Bluhmki E, von Kummer R, et al. ECASS, ATLANTIS, NINDS and EPITHET rt-PA Study Group. Time to treatment with intravenous alteplase and outcome in stroke: an updated pooled analysis of ECASS, ATLANTIS, NINDS, and EPITHET trials. Lancet 2010;375:1695–1703.

[57] Walter S, Kostopoulos P, Haass PA, et al. Diagnosis and treatment of patients with stroke in a mobile stroke unit versus in hospital: a randomised controlled trial. Lancet 2012;11(5):397-404.

[58] IL Katzan, MD Hammer, ED Hixson, et al. Utilization of intravenous tissue plasminogen activator for acute ischemic stroke. Arch Neurol, 61 (2004), pp. 346-350.

[59] Mazighi M, Chaudhry SA, Ribo M, et al. Impact of onset-to-reperfusion time on stroke mortality: a collaborative pooled analysis. Circulation 2013;127:1980–1985.